\documentclass[pdftex,5p,twocolumn]{elsarticle}
\usepackage{txfonts}
\usepackage{natbib}
\bibpunct{(}{)}{;}{a}{}{,}
\usepackage{indentfirst} 
\usepackage{amssymb}
\usepackage{array}
\usepackage{verbatim}

\journal{Icarus}

\begin{document}

\begin{frontmatter}

\title{The quasi-universality of chondrule size as a constraint for chondrule formation models}


\author[cita]{Emmanuel Jacquet\corref{cor1}}
\ead{ejacquet@cita.utoronto.ca}
\cortext[cor1]{Corresponding author}

\address[cita]{Canadian Institute for Theoretical Astrophysics, University of Toronto, 60 St Georges Street, Toronto, ON M5S 3H8, Canada}

\begin{abstract}
Primitive meteorites are dominated by millimeter-size silicate spherules called chondrules. The nature of the high-temperature events that produced them in the early solar system remains enigmatic. Beside their thermal history, one important clue is provided by their size which shows remarkably little variation (less than a factor of 6 for the mean chondrule radius of most chondrites) despite the extensive range of ages and heliocentric distances sampled. It is however unclear whether chondrule size is due to the chondrule melting process itself, or has been simply inherited from the precursor material, or yet results from some sorting process. I examine these different possibilities in terms of their analytical size predictions. Unless the chondrule-forming ``window'' was very narrow, radial sorting can be excluded as size-determining processes because of the large variations it would predict. Molten planetesimal collision or impact melting models, which derive chondrules from the fragmentation of larger melt bodies, would likewise predict too much size variability by themselves; more generally any size modification during chondrule formation is limited in extent by evidence from compound chondrules and the considerable compositional variability of chondrules. Turbulent concentration would predict a low size variability but lack of evidence of any accretion bias in carbonaceous chondrites may be difficult to reconcile with any form of local sorting upon agglomeration. Growth by sticking (especially if bouncing-limited) of aggregates as chondrule precursors would yield limited variations of their final radius in space and time, and would be consistent with the relatively similar size of other chondrite components such as refractory inclusions. This suggests that the chondrule-melting process(es) simply melted such nebular aggregates with little modification of mass.
 \end{abstract}

\begin{keyword}
Meteorites \sep Solar nebula \sep Cosmochemistry \sep Disk \sep Accretion
\end{keyword}

\end{frontmatter}

\section{Introduction}
Primitive meteorites, or \textit{chondrites}, bear witness to the birth of the solar system, 4.57 billion years ago, when the infant Sun was surrounded by a gaseous protoplanetary disk. Beside the \textit{refractory inclusions}---the earliest solids of the solar system \citep[e.g.][]{MacPherson2005}---, chondrites are mostly composed of millimeter-size silicate spherules called \textit{chondrules} \citep{ConnollyDesch2004}. They appear to result from the solidification of molten droplets following  short ($<$ hours or days at most) high-temperature events \citep[e.g.][]{Hewinsetal2005} which must have occurred repeatedly during the evolution of our protoplanetary disk \citep{Jones1996}. Indeed their estimated ages vary in a time span of 0-3 Ma after the formation of refractory inclusions \citep{Villeneuveetal2009,Connellyetal2012}.   

  Despite their ubiquity, the formation mechanism of chondrules, presumably a prominent process in the protoplanetary disk, is still heavily controversial. ``Planetary'' scenarios currently investigated involve impact-induced melting, similar to those invoked for crystalline lunar spherules \citep{Symesetal1998, Sears2005}, or the collision between already molten 
 planetesimals \citep{SandersScott2012,Asphaugetal2011}. Objections against such scenarios include the lack of correlation of chondrules with other expected impact effects, their essentially chondritic bulk composition, their old but variable ages, etc. (see e.g. \citet{Tayloretal1983}). In the last decades, attention has thus focused on ``nebular'' scenarios, where chondrules are interpreted as the products of flash-heating of nebular precursors (e.g. ``dustballs''). While the X-wind model flinging chondrules produced at the inner disk edge outward \citep{Shuetal2001,Hu2010} appears to have fallen out of favor \citep{Deschetal2010}, formation by shock waves, due either to gravitational instabilities \citep{BossDurisen2005} or eccentric planetary embryos \citep{Morrisetal2012} is still a leading contender, with formation in short circuits in magnetohydrodynamical turbulence \citep{McNallyetal2013} or in disk winds \citep{SalmeronIreland2012} being also more recently considered. A serious drawback of these nebular scenarios, though, is the observed retention by chondrules of significant amounts of moderately volatile elements such as Na \citep{Alexanderetal2008,Hewinsetal2012} suggestive, unless chondrules cooled in tens of seconds \citep[e.g.][]{Rubin2000}, of high partial pressures of these in chondrule-forming regions, possibly because of high concentrations of the partially evaporating chondrules themselves---in any case difficult to reconcile with current disk models. Chondrule formation is obviously not a settled issue, nor can the above do justice to all proposed ideas---we refer the interested reader to \citet{Boss1996,Jonesetal2000,DeschConnolly2002,Krotetal2009,Deschetal2012}.

  While most studies have concentrated on the thermal history of chondrules, their sizes also constitute an important constraint. Beyond the absolute scale ($\sim$0.1-1 mm), a striking property is its quasi-universality. Indeed, not only do chondrule sizes vary little in single meteorites (being mostly within a factor of 2 of the mean\footnote{But rare micro- ($\lesssim 10 \mu$m; \citet{Krotetal1997micro}
) and macrochondrules  ($\gtrsim$3 mm; \citet{WeyrauchBischoff2012}) do occur.}; 
 \citet{KingKing1978,Eisenhour1996,Kuebleretal1999, NelsonRubin2002}), but the \textit{mean chondrule size} of individual chondrites spans a limited range of less than a factor of 6 accross all chemical groups (setting aside the CH (10-45 $\mu$m) and CB (2.5 mm for the CB$_a$ subgroup) 
 chondrites, as these petrographically very distinctive chondrites likely had anomalous geneses \citep{Krotetal2005}
). There is moreover little systematic behavior, e.g. as to carbonaceous versus noncarbonaceous chondrites \citep{Benoitetal1999,ScottKrot2003}. This is especially striking given the wide range of chondrule ages or the range of reservoirs that seems required by petrographic specificities of chondrules in different chondrite groups \citep{Jones2012}, from which order-of-magnitude variations of many potentially controlling astrophysical parameters (e.g. density, turbulence etc.)
 could be expected. Whichever process determined chondrule size was thus remarkably insensitive to these variations.   

  The quasi-universality of chondrule size should thus be an important discriminant among different chondrule-forming theories. However, it is \textit{a priori} unclear whether chondrule size was acquired during the chondrule-melting event itself \citep[e.g.][]{Benoitetal1999,Kadonoetal2008,Asphaugetal2011}, or was simply inherited from the precursor \citep{Sekiya1997,Zsometal2010}, or yet was a result of some sorting process \citep{Cuzzietal2001}
. In this paper, I thus examine the effects on chondrule size of the different possible stages in the chondrule cycle, and in particular focus on whether the small \textit{variability} of mean chondrule size among chondrites can be reproduced. Although chondrule size has been previously addressed as part of specific chondrule- or chondrite-forming scenario developments \citep[e.g.][]{Cuzzietal2001,SusaNakamoto2002,MiuraNakamoto2005,Asphaugetal2011}, this is the first attempt at a comprehensive theoretical examination of this question. After a presentation of the overall ``philosophy'' and notations in Section \ref{Models}, I present and discuss size predictions of processes related to precursor growth (Section \ref{growth}), melting (Section \ref{Chondrule formation}) and transport (Section \ref{Sorting}). In Section \ref{Conclusion}, I summarize and conclude.

\section{Generalities}
\label{Models}

\begin{figure}
\resizebox{\hsize}{!}{
\includegraphics{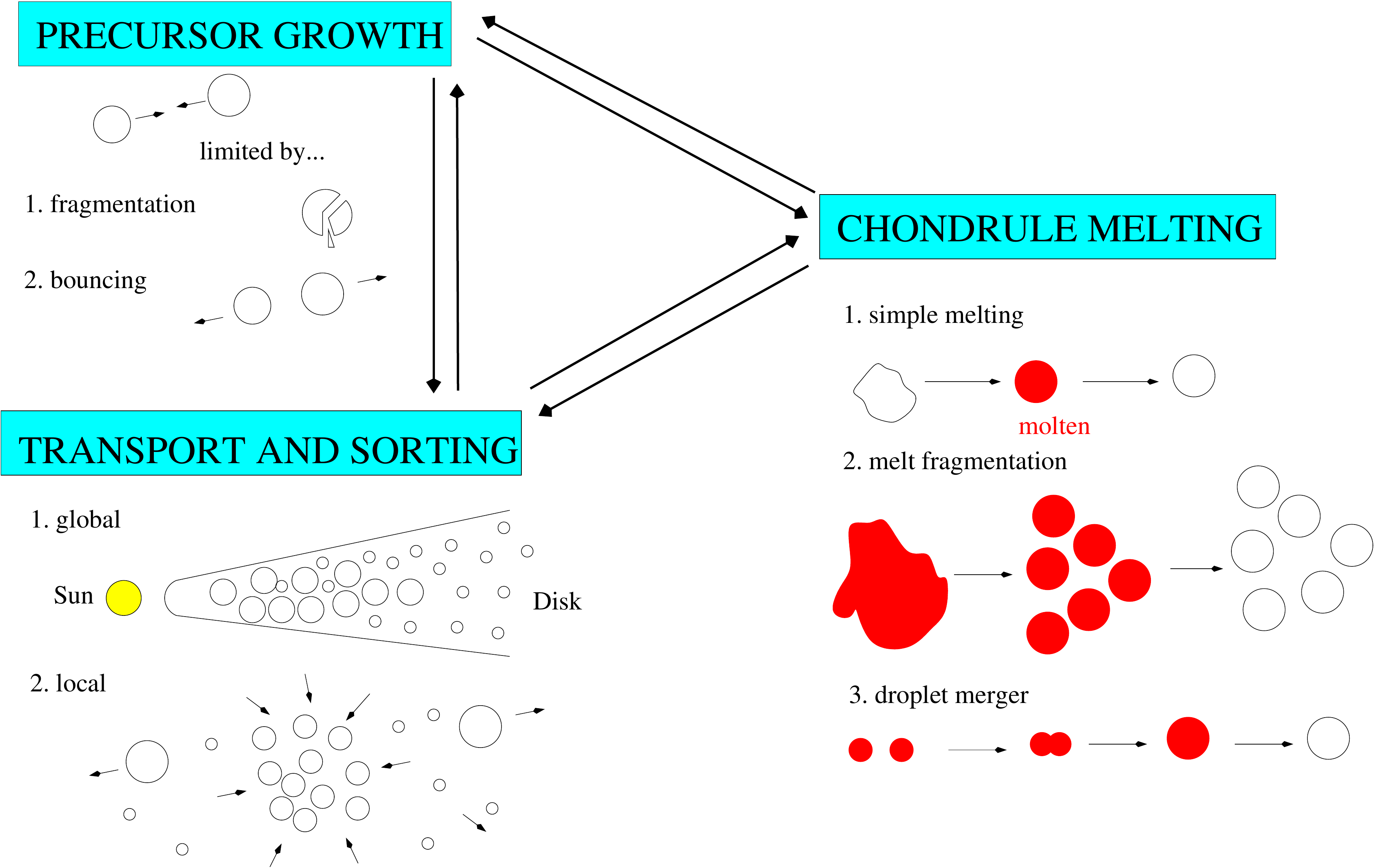}
}
\caption{Sketch of the different envisioned stages in the ``lifecycle'' of chondrules prior to accretion. The arrows indicate that chondrules may have undergone these schematic stages in various sequences. 
 In the chondrule melting phase, a red color marks a molten state. Note that it is not implied that each depicted stage occurred; in fact the relevance of any individual process depends on the correct chondrule-forming scenario.}
\label{life}
\end{figure}

  Schematically, the ``lifecycle'' of a chondrule in the protoplanetary disk, prior to incorporation in a chondrite, involves three broad categories of processes:
\begin{itemize}
\item[(i)] The \textit{growth of chondrule precursors}, limited by fragmentation or bouncing (Section \ref{growth}).
\item[(ii)] \textit{Chondrule formation} proper, which may involve simple melting of preexisting solids, fragmentation of larger melt bodies and/or coagulation of smaller ones (Section \ref{Chondrule formation}). 
\item[(iii)] The \textit{transport} of chondrules/chondrule precursors in the gas and possible associated sorting, either globally (disk-wide) or locally (Section \ref{Sorting}).
\end{itemize}
 These are depicted in Fig. \ref{life}. Depending on the appropriate chondrule formation scenario, a given chondrule/chondrule precursor may have undergone part or all these different stages in various, possibly repeatable sequences until its incorporation in a chondrite. For example, a particular chondrule precursor may have first grown, then melted in a chondrule-forming event, have undergone aerodynamic transport, be remelted in a second chondrule-forming event, before another transport phase and accretion in a planetesimal.

  Each of these stages may have left a ``fingerprint'' \citep{CuzziWeidenschilling2006} on the size distribution of the final chondrules, so in general the shape of the size distribution will result from a complex superposition of several processes
. Nevertheless, the well-defined peaks in the observed chondrule size distributions in chondrites \citep[e.g.][]{Teitleretal2010} are unlikely to be the coincidental result of several processes so that the \textit{mean size} of chondrules in a given chondrite should be essentially traceable to one single process. 
Since, as already emphasized, chondrule mean size does not vary much among different chondrite groups, it would appear that essentially \textit{one} single process determined that chondrule mean size for \textit{all} chondrites. This does not exclude that chondrules in different chondrite groups may have undergone qualitatively different mechanisms affecting chondrule size, but these would be order unity effects.  

  The purpose here is to seek the process which determined the typical size of chondrules in the different chondrites, which I will refer to as the \textit{size-determining process} for short. In the upcoming sections, I will discuss processes relevant to the above stages and express the chondrule size they predict by themselves with simple analytical formulas (including some already available in the literature). As argued above, I will mostly not address the whole size distribution information, which, although richer than the mere datum of the mean size, is \textit{a priori} less diagnostic of single processes, and obviously less straightforward to obtain (let alone analytically) for any given mechanism (see however \citet{Cuzzietal2001,Kadonoetal2008}); in the long run, however, this aspect will have to be investigated for any size prediction model to be considered complete. 

  In evaluating these size predictions, at question will not be that much whether these models can account for the observed chondrule size for \textit{some} plausible disk parameters---we shall see that adequate normalizations can almost always be found---, but whether they can account for the small \textit{variability} of mean chondrule size in chondrites. It must be kept in mind that I am only evaluating whether these processes determined chondrule size, not whether they happened at all. Therefore, even if a given process cannot explain chondrule size, it will not be necessarily ruled out as a part of the lifecycle of chondrules, but will require the existence of another mechanism, either before or after it
, which can. 

 In the following, I will denote by $\rho_s$ and $a$ the chondrule (or chondrule precursor) internal density and radius respectively. The gas surface density is $\Sigma$. Wherever the local gas density $\rho$ will be needed, I will use the midplane value of a vertically isothermal disk in hydrostatic equilibrium, i.e. $\rho=\Sigma/\left(\sqrt{2\pi}H\right)$, with the pressure scale height $H=c_s/\Omega$, where $c_s=\sqrt{k_BT/m}$ and $\Omega$ stand for the sound speed (with $T$ the temperature and $m=3.9\times 10^{-27}$kg the mean molecular mass) and Keplerian angular velocity, respectively. Gas drag in the Epstein regime is characterized by the stopping time 
\citep{Epstein1924}:  
\begin{equation}
\tau = \sqrt{\frac{\pi}{8}}\frac{\rho_sa}{\rho c_s}.
\end{equation}
The gas is finally assumed to be turbulent, with $\alpha$ the local standard \citet{ShakuraSunyaev1973} parameter.

\section{Chondrule precursor growth}
\label{growth}

  Chondrules are widely believed to result from the melting of preexisting solids called ``precursors''. In that case, a first possibility is that chondrule size is inherited from them and thus dictated by the primary coagulation process of solid grains \citep{Sekiya1997,Zsometal2010} 
\citep{Chokshietal1993,DominikTielens1997,BlumWurm2008, Guettleretal2010}. In inner regions of the protoplanetary disk, growth is likely to be limited by a velocity threshold for sticking rather than by inward drift \citep{Birnstieletal2012}.

  For many limiting processes \citep[e.g.][]{Guettleretal2010,Beitzetal2011}, the critical sticking velocity can be cast in the form:
\begin{equation}
\label{vstick}
\Delta v=v_{\rm ref}\left(\frac{m_p}{m_\ast}\right)^{-\delta}
\end{equation}
with $m_p$ the aggregate mass and $m_\ast$, $v_{\rm ref}$, $\delta$ fixed parameters.

 In a turbulent disk, the particle-particle velocity may be approximated by:
\begin{equation}
\label{vpp}
\Delta v=\mathrm{max}\left(\Delta\tau\frac{||\nabla P||}{\rho},\sqrt{\alpha}c_s\:\mathrm{min}\left(\mathrm{Re}^{1/4}\Omega\Delta\tau,\sqrt{3\Omega\tau}\right)\right),
\end{equation}
where $\Delta\tau$ is the absolute difference in stopping time (due to finite dispersion (in size and shape), assumed here to be of order $\tau$) and $||...||$ is the euclidean norm. The first contribution is meant to be that of the background pressure gradient (for $\tau\ll\Omega^{-1}$; \citet{YoudinGoodman2005}), while the second one is that of turbulent fluctuations (using approximations in section 3.4 of \citet{OrmelCuzzi2007} for $\tau$ smaller and larger than the Kolmogorov timescale $\mathrm{Re}^{-1/2}\Omega^{-1}$, respectively), where 
\begin{equation}
\mathrm{Re}=\frac{\alpha c_s H}{\nu_{\rm mol}}=2\sqrt{2}\frac{\Sigma\sigma_{H_2}\alpha}{m}
\end{equation}
is the Reynolds number, with $\nu_{\rm mol}$ the molecular kinematic velocity of the gas and $\sigma_{H_2}=5.7\times 10^{-20}\:\mathrm{m}^2$ the collisional cross section of H$_2$ \citep{Cuzzietal2001}.

  The size resulting from growth by sticking is then obtained by equating the particle-particle velocity (\ref{vpp}) and the critical sticking velocity (\ref{vstick}). Expressed in term of the compact-equivalent radius $a_{\rm comp}\equiv a\phi^{1/3}$ with $\phi$ the volume filling factor, this size is given by\footnote{If $\Delta\tau\ll\tau$ (unlike what I assume here), a significant error will be incurred if the stopping time corresponding to the second line of equation (\ref{acomp general}) is not explicitly constrained to be smaller than $\mathrm{Re}^{-1/2}\Omega^{-1}$.
}:
\begin{eqnarray}
\label{acomp general}
a_{\rm comp}=\mathrm{min}\Bigg(\left(\frac{2}{\pi}\frac{\tau}{\Delta\tau}\frac{v_{\rm ref}}{\rho_s^{1+\delta}\phi^{2/3}c_s}\left(\frac{3m_\ast}{4\pi}\right)^{\delta}\frac{\Sigma}{H ||\nabla \mathrm{ln}P||}\right)^{1/(1+3\delta)},\nonumber\\
\mathrm{max}\Bigg(\left(\frac{\tau}{\Delta\tau}\frac{2^{5/8}v_{\rm ref}}{\pi\rho_s^{1+\delta}\phi^{2/3}c_s}\left(\frac{3m_\ast}{4\pi}\right)^{\delta}\left(\frac{\Sigma}{\alpha}\right)^{3/4}\left(\frac{m}{\sigma_{H_2}}\right)^{1/4}\right)^{1/(1+3\delta)},\nonumber\\
\left(\left(\frac{v_{\rm ref}}{c_s}\right)^2\left(\frac{3m_\ast}{4\pi}\right)^{2\delta}\frac{2}{3\pi\rho_s^{1+2\delta}\phi^{2/3}}\frac{\Sigma}{\alpha}\right)^{1/(1+6\delta)}\Bigg)\Bigg)
\end{eqnarray}
with $\rho_s$ meant to be the compact density (set at $3\times 10^3\:\mathrm{kg/m^3}$).

With this general formula at hand, we now examine the specific cases of fragmentation- and bouncing-limited regimes.

\subsection{Fragmentation-limited growth}
\label{Fragmentation}

  Fragmentation is generally modeled with a size-independent ($\delta=0$) velocity threshold $v_{\rm ref}\approx1-10\:\mathrm{m/s}$ \citep[e.g.][]{Guettleretal2010,Birnstieletal2012}. The stopping time would be in the inertial range of the Kolmogorov cascade, yielding:
\begin{eqnarray}
\label{a fragmentation}
a_{\rm comp}=\frac{2}{3\pi\rho_s\phi^{2/3}}\frac{\Sigma}{\alpha}\left(\frac{v_{\rm ref}}{c_s}\right)^2=0.7\:\mathrm{mm}
\left(\frac{\Sigma}{10^4\:\rm kg/m^2}\right)\nonumber\\\left(\frac{10^{-3}}{\alpha}\right)\left(\frac{v_{\rm ref}}{1\:\rm m/s}\right)^2\left(\frac{300\:K}{T}\right)\phi^{-2/3}.
\end{eqnarray}
Although protoplanetary disk physics are not well-understood, for evolution times of a few Ma, and heliocentric distances spanning say 1-5 AU at least, variations of $\Sigma$ and $\alpha$ by \textit{at least} 1-2 order of magnitude \textit{each} may be robustly expected \citep[e.g.][]{Hayashi1981,Desch2007,Chambers2009,Turneretal2010,YangCiesla2012,Flocketal2011}. Thus, unless chondrule formation took place on a temporally and spatially very narrow window
, it appears that a fixed fragmentation threshold would predict too much variation of chondrule size by itself to be the size-determining process.

  There is however evidence that fragmentation velocity may depend on size. Experiments by \citet{Beitzetal2011} suggest $\delta=0.158$ and $m_\ast=3.67\times10^4$ kg for $v_{\rm ref}=$1 cm/s (see also \citet{Windmarketal2012a}). In that case, the size (still corresponding to a stopping time in the inertial range) becomes:
\begin{equation}
a_{\rm comp}=0.6\:\mathrm{mm}\left(\frac{\Sigma}{10^4\:\rm{kg/m^2}}\frac{10^{-3}}{\alpha}\frac{300\:\rm K}{T}\right)^{0.51}\phi^{-0.32}.
\end{equation}
With a reduced dependence on $\Sigma/\alpha$ (now to the 0.51 power), this could satisfy the chondrule size variability constraint.

\subsection{Bouncing-limited growth}
\label{Bouncing}

 Particles may bounce at velocities much lower than the fragmentation limit \citep{Zsometal2010}, which may thus set the real limit to growth\footnote{Although some fragmentation in the upper layers is still needed to replenish the micron-size grains there \citep{DullemondDominik2005,
Braueretal2008,
Zsometal2011}.}. Although numerical simulations of aggregate collisions have hitherto failed to reveal this ``bouncing barrier'' for porous aggregates---the expected products of initial hit-and-stick growth---\citep[e.g.][]{SeizingerKley2013}, \citet{Kotheetal2013} experimentally found a bouncing velocity parameterized by $\delta = 3/4$ and $m_\ast=10^{-7.3}$ kg for $v_{\rm ref}=$ 1 cm/s, yielding a limiting size of:
\begin{equation}
\label{a Kothe}
\begin{array}{rrrr}
 a_{\rm comp}  =\mathrm{min}\Bigg(0.3\:\mathrm{mm} \left(\frac{\Sigma}{10^5\:\rm kg/m^2}\frac{\Delta\tau/\tau}{H ||\nabla \mathrm{ln}P||}\right)^{4/13}
\left(\frac{0.1}{\phi}\right)^{8/39}&\nonumber\\
 \left(\frac{300\:\rm K}{T}\right)^{2/13},&\nonumber\\
\mathrm{max}\Bigg( 0.1\:\mathrm{mm}\left(\frac{\Sigma}{10^5\:\rm kg/m^2}\frac{10^{-3}}{\alpha}\right)^{3/13}
\left(\frac{0.1}{\phi}\right)^{8/39}
 \left(\frac{300\:\rm K}{T}\right)^{2/13}\left(\frac{\tau}{\Delta\tau}\right)^{4/13},&\nonumber\\
  0.08\:\mathrm{mm}\left(\frac{\Sigma}{10^5\:\rm kg/m^2}\frac{10^{-3}}{\alpha}\frac{300\:\rm K}{T}\right)^{2/11}\left(\frac{0.1}{\phi}\right)^{4/33}\Bigg)\Bigg)&
\end{array}
\end{equation}
So depending on the relevant regime, we obtain dependences in $\Sigma^{4/13}$, $(\Sigma/\alpha)^{3/13}$ or $(\Sigma/\alpha)^{2/11}$. To get a sense of the sensitivity of that result to the uncertainties of the bouncing barrier, we note that for theoretical thresholds discussed by \citet{Guettleretal2010}, the exponents would become respectively 2/5, 3/10 and 1/4 for ``hit-and-stick growth'' (``S1''; $\delta=1/2$) and 6/11, 9/22 and 3/8 for ``sticking with surface effects'' (``S2''; $\delta=5/18$). These are weak dependences (a range of 0.18-0.41). This is because the positive and negative dependences of the relative velocity (equation (\ref{vpp})) and the sticking velocity (equation (\ref{vstick})), respectively, add up when these are equated to each other, hence a weak dependence of size on disk parameters after solving for it. 

  The result also weakly depends on the porosity of the aggregates, so long the initial increase due to fractal growth has been checked e.g. by compression at moderate collision speeds \citep[e.g.][]{Ormeletal2007,Zsometal2010}. However, porosity evolution models by \citet{Okuzumietal2009}, whose numerical experiments show greater porosity increases in unequal-sized collisions than modeled by \citet{Ormeletal2007}, lead to spectacular decreases of $\phi$ by more than 3 orders of magnitude \citep{Okuzumietal2012}, and the bouncing barrier (not included) may be less stringent for such aggregates \citep{Kotheetal2013}. However, very low $\phi$ are probably unrealistic---in fact, pre-compaction estimates for fine-grained rims around Allende chondrules, presumably accreted in the disk, are 20-30 \% \citep{Blandetal2011}---, because not all monomers were micron-sized
, which would have placed lower bounds on the overall density and stopping times (and possibly affected sticking properties \citep{Ormeletal2008,Beitzetal2012}).  Indeed, the compositional variability of chondrules \citep{HezelPalme2007} indicate that chondrule precursors contained coarse grains ($\gtrsim 10-100\:\mu$m), e.g. refractory inclusions or earlier chondrule debris, as sometimes evidenced by relict grains \citep{Jones1996}. The precursors may have been analogous to agglomeratic olivine objects or amoeboid olivine aggregates found in chondrites \citep{Ruzickaetal2011, Ruzickaetal2012AOA}, both of probable nebular origin. The overall similarity in size (within a factor of a few) of the latter 
 with chondrules would be also consistent with a derivation of chondrule size from nebular aggregates, although the systematically smaller size of refractory inclusions would have yet to be explained (e.g. different available time, monomer size, temperature...). 

  In the above, we have considered only \textit{typical} collision speeds to estimate the final size, but especially for the turbulent contribution, a \textit{distribution} of velocities would be more realistic, and particles may grow beyond the bouncing barrier calculated above through a series of low-probability, low-velocity collisions \citep[e.g.][]{Windmarketal2012,Garaudetal2013}. While meter-size particles would be very rare
, the effective ``bottleneck'' of the coagulation may be increased above our nominal bouncing barrier, and would be asymptotically determined by a balance between sticking and fragmentation probabilities \citep{Windmarketal2012}
. However, the simulations of \citet{Windmarketal2012} and \citet{Garaudetal2013}, both finding a shift of one order of magnitude, did not take into account the decrease of the bouncing speed with increasing size (not to mention possible increased fragmentability), and their assumption of a Maxwellian distribution may overestimate the low-velocity tail (susceptible to sticking) of the true distribution at the expense of the high-velocity one (susceptible to fragmentation) \citep{PanPadoan2013}, both of which effects would lead not only to a decrease of the asymptotic limit, but also to a steep increase of the time needed for growth (inversely proportional to the sticking probability) beyond our nominal bouncing barrier. Still, the ``lucky'' sticking events could account for the large-size tail of chondrule size distributions.

\section{Chondrule melting}
\label{Chondrule formation}

  Certainly, the cornerstone of the lifecycle of chondrules is the chondrule-melting mechanism itself. Although it might have merely melted preexisting aggregates, molten bodies could have fragmented or coagulated, hence a size modification between the precursors and the chondrules, as I envision now.

\subsection{Simple melting of precursor}

  I call ``simple melting'' a scenario where chondrule size (or more properly, mass) is inherited from the precursors, without significant mergers or disruptions. This is generally implied in the conventional ``flash-heating'' picture of many ``nebular'' scenarios, e.g. shockwave \citep{Deschetal2005}, X-wind \citep{Shuetal2001}, lightning \citep{DeschCuzzi2000} or short circuits in magnetorotational turbulence \citep{McNallyetal2013}.

  Simple melting scenarios would not be expected to leave any fingerprint on chondrule size distribution
, unless the mechanism responsible somehow preferentially processed solids from one size bin, or destroyed the solids from other size bins. The model closest to the former situation known to me seems to be the recent \citet{SalmeronIreland2012} scenario of chondrule formation in disk winds, where only particles of a specific size bin are small enough to be first entrained upward and big enough to then fall back (because of dust accretion), whereupon gas drag leads to melting.
For the time being, 
 no expression of this preferred size as a function of disk parameters is available
. As to selective destruction, investigators of the shockwave models have also proposed that large droplets would get disrupted by the ram pressure appearing upon crossing the shock front \citep{SusaNakamoto2002} while tiny ones would be evaporated \citep{MiuraNakamoto2005}. But the two resulting size cutoffs would not in general coincide even within a factor of a few\footnote{They would also likely be quite variable. For a velocity jump $v_{\rm pg}$, the maximum size is given by \citep{SusaNakamoto2002,Kadonoetal2008}:
\begin{eqnarray}
a &=& \frac{2\gamma\mathrm{We}}{\rho_{\rm post}v_{pg}^2}\approx \frac{2\gamma\mathrm{We}\rho_{\rm pre}}{\rho_{\rm post}^2c_{s,\rm post}^2}\sim \frac{\gamma}{\rho_{\rm pre}c_{s, \rm post}^2}\nonumber\\
&=& 6\:\mathrm{mm}\left(\frac{\gamma}{0.4\:\rm N/m}\right)\left(\frac{10^{-5}\:\rm kg/m^3}{\rho_{\rm pre}}\right)\left(\frac{2000\: \rm K}{T_{\rm post}}\right)
\end{eqnarray}
with $\mathrm{We}\approx 6$ the critical Weber number and the ``pre'' and ``post'' subscripts referring to the pre- and post-shock regions respectively (for the second equality, I have used $\rho_{\rm pre}v_{pg}^2\approx\rho_{\rm post}c_{s,\rm post}^2$ from momentum flux conservation
). 
}
 so that the typical size of chondrules would rather be inherited from precursors than set by the shock event.

\subsection{Melt fragmentation}
\label{Top-down}

  Other scenarios envision chondrules as fragments of larger melt bodies. Two illustrative models will be considered here: the molten planetesimal collision (or ``splashing'') scenario \citep[e.g.][]{Asphaugetal2011,SandersScott2012} and an hypervelocity impact plume \citep[e.g.][]{Symesetal1998}.

  In the splashing scenario of \citet{Asphaugetal2011}, melt is already present in the planetesimal before collision as a result of $^{26}$Al decay. Moderate-velocity collision eject sheets of this melt which fragment as pressure unloads until the  Laplace pressure $2\gamma/a$ (with $\gamma$ the surface tension) of the droplets essentially balances the original (hydrostatic) pressure in the source planetesimal prior to impact, resulting in a typical radius of
\begin{eqnarray}
\label{a splash}
a &=& \frac{2\gamma}{GE\rho_p^2R_p^2}\\
&=& 0.1\:\mathrm{mm}\left(\frac{\gamma}{0.4\:\rm N/m}\right)\left(\frac{0.5}{E}\right)\left(\frac{3\times 10^3\:\rm kg/m^3}{\rho_p}\frac{5\:\rm km}{R_p}\right)^2,\nonumber
\end{eqnarray}
with $R_p$, $\rho_p$ the radius and density of the source planetesimal, respectively, and $E$ an efficiency factor 
 \citep{Asphaugetal2011}.  To be the size-determining process, the splashing scenario would require the colliding bodies to have a very narrow size distribution around $\sim$10 km (within a factor of 3)
, which is not borne out by that of the asteroid main belt or earlier modeled stages thereof \citep{Davisetal2002,Morbidellietal2009,Weidenschilling2011}.

  For impact velocities $\gtrsim$ 3 km/s \citep{Stoeffleretal1991}, 
melt can be produced by the impact itself \citep{Symesetal1998,Benoitetal1999}. The initial fragmentation of the liquid yields clumps of radius (\citet{MeloshVickery1991}, see also \citet{JohnsonMelosh2014}):
\begin{eqnarray}
\label{a tektite}
a &=& \left(\frac{20\gamma}{\rho_s}\right)^{1/3}\left(\frac{R_i}{v_i}\right)^{2/3}\nonumber\\
&=& 1\:\mathrm{cm}\left(\frac{\gamma}{0.4\:\rm N/m}\right)^{1/3}\left(\frac{R_i}{0.1\:\rm km}\right)^{2/3}\left(\frac{5\:\rm km/s}{v_i}\right)^{2/3}
\end{eqnarray}
with $R_i$, $v_i$ the impactor radius and velocity, respectively. For $v_i\gtrsim 10-15\:\rm km/s$, partial vaporization of the target, in addition to melting, entails a competition between aerodynamic forces and surface tension of the droplets, yielding a minimum size of liquid droplets of \citep{MeloshVickery1991}
\begin{equation}
\label{a Melosh}
a=0.3\:\mathrm{mm}\left(\frac{\gamma}{0.4\:\rm N/m}\right)^{1/2}\left(\frac{R_i}{1\:\rm km}\right)^{1/2}\left(\frac{15\:\rm km/s}{v_i}\right).
\end{equation}
While the dependence on impactor radius is weak
, that scenario would predict the correlated existence of larger ($\sim$cm-size) glassy objects (equation \ref{a tektite}), contrary to observations (although a few impact melts occur in chondrite breccias \citep[e.g.][]{Keiletal1980}). As a general problem with melt fragmentation scenarios, large chondrule-textured objects are persistently lacking \citep{Tayloretal1983}---the largest chondrule known to date being the 5 cm-diameter ``Golfball'' in the Gunlock L3 chondrite \citep{Prinzetal1988}. I however note that the large CB$_a$ and small  CH chondrite chondrules would match the larger size variations expected from impact, which would be consistent with the generally agreed impact-induced formation of these objects \citep{Krotetal2005} which, it must be reminded, are very different from mainstream chondrules. 

\subsection{Droplet mergers}
\label{Bottom-up}

  One could envision that currently observed chondrules result from the mergers of smaller liquid droplets. Ignoring fragmentation, the growth of chondrules would be given by
\begin{equation}
\frac{\mathrm{d}}{\mathrm{dt}}\left(\frac{4\pi}{3}\rho_sa^3\right)=4\pi a^2 \rho_d \Delta v,
\end{equation}
with $\rho_d$ the droplet mass density 
and $\Delta v$ a typical droplet-droplet velocity at time $t$. The final radius (if, ex hypothesi, the initial one can be neglected) will then be: 
\begin{eqnarray}
\label{Delta a}
a &=& \frac{\rho_d}{\rho_s}\Delta v t_m\nonumber\\
&=& 0.2\:\mathrm{mm}\left(\frac{\rho_d}{10^{-4}\:\mathrm{kg/m^3}}\right)\left(\frac{\Delta v}{1\:\rm m/s}\right)\left(\frac{t_m}{10\:\rm h}\right),
\end{eqnarray}
with $t_m$ the time during which droplet mergers took place. 

  To my knowledge, no chondrule formation model has predicted chondrule sizes this way (which would require very high solid densities). 
One may however empirically evaluate the importance of collisions from the abundance of compound chondrules, i.e. chondrules fused together \citep{GoodingKeil1981,Wassonetal1995,Cieslaetal2004,AkakiNakamura2005}, whose formation differs from the mergers envisioned above 
 only in that the fused components did not have time to relax to one spherical object. The average compound chondrule fraction in ordinary chondrites is only 4 \% \citep{GoodingKeil1981} and may be twice higher in CV chondrites \citep{AkakiNakamura2005}, suggesting that collisions were not frequent enough to significantly affect the size distribution of chondrules. 

  However, cooling histories may be conceivable for which the time span where compound droplets did not have time to relax to sphericity was only a small fraction of the time where mergers were complete \citep{AlexanderEbel2012}.
As to this possibility, an important clue is given by the fact that nonporphyritic chondrules (i.e. chondrules which have been most efficiently melted with most crystal nuclei destroyed before cooling) have a higher compound fraction (up to 28 \% according to \citet{GoodingKeil1981}) than their 
 porphyritic counterparts. In \ref{Compound model}, I argue, with a simple modeling of compound chondrule formation, that colliding pairs involving one or two totally molten droplets may have frozen in the compound shape at higher temperatures than already crystallizing ones, which might explain this. Alternatively, nonporphyritic chondrules may have generally formed in distinct environments with higher collision rates \citep{GoodingKeil1981,Cieslaetal2004}. 
In either case, a size difference between nonporphyritic and porphyritic chondrules would be expected if droplet mergers significantly affected chondrule size. 
 Such a difference does exist, with nonporphyritic chondrules being on average bigger than porphyritic ones \citep[e.g.][]{RubinGrossman1987,Rubin1989,NelsonRubin2002,WeyrauchBischoff2012}, 
but is limited, usually within a factor of 2 (and \citet{NelsonRubin2002} even suggest it to be an artifact of preferential fragmentation of large porphyritic chondrules on the parent-body), although droplet densities and/or cooling timescales may have varied by orders of magnitude. This is evidence that mergers had a marginal effect on size in general
.

  The variability of chondrule composition provides an independent general limitation on the importance of mergers, which also pertains to ``melt fragmentation'' scenarios discussed in Section \ref{Top-down} since the large melt bodies to be disrupted, or their precursors, would have had to be produced by the merger of preexisting bodies as well. \citet{HezelPalme2007} showed that no more than 10-100 precursor grains could have contributed to each chondrule given the observed variances, limiting any \textit{radius} change to a factor of a few.
One could argue though that the compositional variability of chondrules in a given chondrite is due to the diversity of their source reservoirs. This is certainly a contribution and the cosmochemical fractionation trends exhibited by chondrules are essentially the same as those of bulk chondrites \citep{GrossmanWasson1983}. 
But if mergers were so numerous as to homogenize chondrule compositions in each chondrule-forming region, one would expect components of compound chondrules to have very similar compositions. 
While bulk chemical data are currently lacking for these objects and are certainly desirable to settle the matter, differences in modal mineralogy and mineral chemistry in many of them \citep{Wassonetal1995,AkakiNakamura2005} 
make this fairly unlikely. Another problem is the existence of a sizable proportion ($>$10 \%) of chondrules with anomalous rare earth element (REE) abundance patterns \citep{MisawaNakamura1988,Packetal2004} 
presumably inherited from refractory precursors \citep[see e.g.][]{Boynton1989}. Refractory inclusions, formed during $\lesssim 10^5$ years after the building of the protoplanetary disk \citep[e.g.][]{Bizzarroetal2004}, presumably within a few AUs of the Sun, would be rapidly mixed together\footnote{For example, the turbulent diffusion timescale $R^2/\left(\alpha c_s^2\Omega^{-1}\right)$ is about $5\times 10^4$ years for R = 1 AU, T = 1500 K, $\alpha=10^{-3}$.}
, before the formation of most chondrules. One would thus not expect any chondrule-forming reservoir \textit{as a whole} to show such anomalous REE patterns 
 and they have not been observed at the scale of bulk chondrites  \citep[e.g.][]{Evensenetal1978,Boynton1984}. 
This argues against reservoir-scale compositional homogenization of grains. 

  Petrographical and compositional evidence from chondrules thus appears inconsistent with significant (order-of-magnitude) size modification during chondrule formation. 
More subordinate size modifications might however explain part 
of the variations of chondrule size (see e.g. \citet{Rubin2010}), and perhaps the systematic size difference with refractory inclusions.

\section{Transport and sorting}
\label{Sorting}

  The age range of chondrules in single chondrites \citep{Villeneuveetal2009,Connellyetal2012} suggests that they spent a few million years as free-floating objects in the protoplanetary disk. Aerodynamic transport may then have resulted in size sorting of chondrules or their precursors, either on a global or a local scale, as I now examine.

\subsection{Global sorting}
\label{Global sorting}


  \citet{Jacquetetal2012S} showed that the dynamic response of particles embedded in the gaseous disk was essentially governed by the ``gas-grain decoupling parameter'' $S\equiv \Omega\tau/\alpha$. For $S\ll 1$, particles are well-coupled to the gas, while for $S>1$, they tend to settle to the midplane and drift toward the Sun. If chondrules (or chondrule precursors) were produced in the inner regions of the disk and subsequently redistributed in outer regions of the disk, their outward transport would essentially be stopped at the $S=1$ line \citep{Jacquetetal2012S}. Thus, at a given heliocentric distance, particles larger than
\begin{equation}
\label{a S=1}
a\approx\frac{2\Sigma\alpha}{\pi\rho_s}=0.2\:\mathrm{mm}\left(\frac{\Sigma}{10^3\:\mathrm{kg/m^2}}\right)\left(\frac{\alpha}{10^{-3}}\right)\left(\frac{3\times 10^3\:\mathrm{kg/m^3}}{\rho_s}\right)
\end{equation}
would have been prevented from reaching this location so that equation (\ref{a S=1}) would provide a maximum cutoff\footnote{We note that, regardless of where chondrules were produced, vertical settling would concentrate particles \textit{larger} than this same size (roughly) at the midplane. Settling and radial segregation could act together to narrow the size distribution around the size given.}. Although $\Sigma$ and $\alpha$ may be anticorrelated, e.g. in a steady disk or a dead zone \citep[e.g.][]{FlemingStone2003, Terquem2008}, this would not prevent order-of-magnitude variations to arise \citep{Jacquetetal2012S}, contrary to observations. Radial transport in the disk is thus unlikely to have been the size-determining process.


  Solids may have been transported in winds rather than through the disk. In the X-wind scenario, the chondrules transported to the planet-forming region (rather than falling back close to the X point or flying toward interstellar space) must have a stopping time comparable to the orbital period at the X point \citep{Shuetal1996}. \citet{Shuetal1996} define in their equation 4 a parameter which I will call $\zeta$ for which they infer a critical value of 0.4, from which the selected size may be expressed as:
\begin{eqnarray}
\label{X wind}
a &=& \frac{\dot{M}^{8/7}(4\pi/\mu_0)^{1/7}}{16\pi\rho_s\zeta \left(GM_\ast\right)^{3/7}\mu_\ast^{2/7}}\nonumber\\
&=& 0.2\:\mathrm{mm}\left(\frac{3\times 10^3\:\mathrm{kg/m^3}}{\rho_s}\right)\left(\frac{0.8\:\rm M_\odot}{M_\ast}\right)^{3/7}\left(\frac{10^{34}\:\mathrm{A m^2}}{\mu_\ast}\right)^{2/7}\nonumber\\
& &\left(\frac{0.4}{\zeta}\right)\left(\frac{\dot{M}}{10^{-7}\:\rm M_\odot/a}\right)^{8/7}
\end{eqnarray} 
with $\dot{M}$ the mass accretion rate from the disk, $M_\ast$ and $\mu_\ast$ the mass and magnetic dipole moment of the star (evaluated here in the ``revealed'' stage of the X-wind). Detailed trajectory calculations by \citet{Hu2010} are consistent with this prediction within factors of order unity. While there would be little spatial variation, the dependence on the accretion rate ($\propto \dot{M}^{8/7}$) would entail a spread over $\geq 1$ order of magnitude, which while possibly marginally reconcilable with observations---although it would predict a correlation between chondrule age and size---, may be greatly exacerbated by radial fluctuations in the position of the launching point \citep{CuzziWeidenschilling2006}. I note that the very role of the X-wind in processing chondrite components has been criticized by \citet{Deschetal2010}. 


When the disk becomes optically thin, photophoresis would become important and may entail radial sorting of the remaining solids according to the product of the density and the thermal conductivity \citep{WurmKrauss2006}. In itself, however, this would induce no \textit{size} sorting, although some size dependence of conductivity might be expected if chondrules are not bare but embedded in dusty aggregates \citep{WurmKrauss2006}. This would however require these dusty aggregates to have already exhibited some chondrule size selection, thus established prior to photophoretic transport.

\subsection{Local sorting}
\label{Local sorting}

  Sorting could alternatively have been \textit{local}, perhaps as a prelude to planetesimal formation. \citet{Cuzzietal2001} proposed that turbulence concentrated particles between eddies, most efficiently for stopping times equal to the Kolmogorov timescale
, corresponding to a radius of\footnote{The numerical factors differ slightly from those of \citet{Cuzzietal2001} because of the typical density I have chosen.}:
\begin{eqnarray}
\label{a TC}
a &=&\frac{2^{1/4}}{\pi\rho_s}\left(\frac{\Sigma m}{\alpha \sigma_{H_2}}\right)^{1/2}\nonumber\\
&=& 0.1\:\mathrm{mm}\left(\frac{\Sigma}{10^4\:\mathrm{kg/m^2}}\frac{10^{-3}}{\alpha}\right)^{1/2}\left(\frac{3\times 10^3\:\rm kg/m^3}{\rho_s}\right).
\end{eqnarray}
The thus concentrated particles could pave the way to planetesimal formation \citep{Cuzzietal2001,Cuzzietal2010} and/or provide the dense environment required for chondrule formation \citep{CuzziAlexander2006}, which would thus account for the prefered chondrule size. Other proposed accretion processes such as the streaming instability \citep{YoudinGoodman2005,Johansenetal2007,Johansenetal2009,BaiStone2010b,Jacquetetal2011b} 
may also be accompanied by some size sorting but this has not been investigated to date.

  The $(\Sigma/\alpha)^{1/2}$ dependence predicted by the turbulent concentration scenario is comparable to that of precursor growth by sticking (if somewhat stronger than for bouncing-limited growth). It is notable that the detailed size distribution predicted by turbulent concentration would match the observed size (or more precisely here, $\rho_sa$) distribution of chondrules in ordinary chondrites \citep{Cuzzietal2001,Teitleretal2010}, although that depends on the assumed pre-sorting distribution. It would also be consistent with 
 rough aerodynamical equivalence of other components \citep{Hughes1978,Hughes1980,SkinnerLeenhouts1993,Kuebleretal1999}. 

  However, this equivalence is far from perfect, with e.g. 
 metal grains 
 \citep{Schneideretal1998,NettlesMcSween2006} or refractory inclusions \citep{Mayetal1999,Hezeletal2008}\footnote{With the exception of type B refractory inclusions in CV chondrites, which are on the contrary \textit{larger} (mm-cm) than neighbouring chondrules.} having generally smaller $\rho_sa$ than the colocated chondrules, and the size difference between porphyritic and nonporphyritic chondrules
, not to mention the dust that formed the matrix. We show in \ref{Nonspherical Epstein} that taking into account nonspherical shapes only decreases the stopping time for a given volume and thus widens the discrepancy. As a result
, the very size-selective turbulent concentration would be expected to introduce a change in the proportions of components accreted relative to those present in the original reservoir: this is what \citet{Jacquetetal2012S} called ``accretion bias''. However, while such sorting could account for metal/silicate fractionation in noncarbonaceous chondrites \citep[e.g.][]{Zandaetal2006}, carbonaceous chondrites \citep[e.g.][]{PalmeJones2005,Ebeletal2008} show little evidence of such an accretion bias, as they display near solar Fe/Si and Mg/Si ratios, and also \textit{super}solar Al/Si ratios---indicating, if anything, an \textit{over}abundance of refractory inclusions 
 rather than an undersampling due to non aerodynamical equivalence. This makes any kind of syn-accretional local sorting difficult to envision for the bulk of chondrite accretion. 

  It is important to note, though, that the complementarity between chondrules and matrix, which control the Mg/Si ratio \citep{HezelPalme2010}, could be ensured in a local sorting scenario if the grains constituting the matrix accreted on the individual chondrules prior to sorting, provided (i) the bulk of the dust did end up captured that way and (ii) the amounts accreted were proportional to chondrule mass, for which there is both empirical \citep{Metzleretal1992} and theoretical \citep{Ormeletal2008} support. In this case, size selection of the dust-coated chondrules (which would amount to a size selection of the embedded chondrules, modulo a constant factor) would not affect the (complementary) chondrule-to-matrix ratio. This would not however explain the overabundance of refractory inclusions in carbonaceous chondrites, unless the size distribution of the former was modified after accretion (similar to the suggestion of \citet{NelsonRubin2002} for some chondrules). Note that these arguments against syn-accretional sorting would not apply to sorting prior to or during \textit{chondrule} formation (as suggested by \citet{Morrisetal2012} for the bow shock model), provided it was temporally distinct from \textit{chondrite} accretion.

\begin{table}
\caption{Summary of chondrule size models. (The reader is referred to the text for the definition of symbols and numerical applications).}
\label{Table size}
\begin{tabular}{c c c}
\hline \hline
Category & Process name & Predicted radius\\
\hline\\[-1.5ex]
 Precursor growth & fragmentation-limited & $\left(\frac{2\left(3m_\ast\right)^{2\delta}\left(v_{\rm ref}/c_s\right)^2}{3\pi(4\pi)^{2\delta}\rho_s^{1+2\delta}\phi^{2/3}}\frac{\Sigma}{\alpha}\right)^{\frac{1}{1+6\delta}}$ 
 \\[1ex]
 & bouncing-limited & Equation (\ref{acomp general})
\\[1ex]
\hline\\[-1.5ex]
Transport & radial sorting & $\frac{2\Sigma\alpha}{\pi\rho_s}$ \\[1ex]
& X wind & $\frac{\zeta^{-1}\dot{M}^{8/7}(4\pi/\mu_0)^{1/7}}{16\pi\rho_s\left(GM_\ast\right)^{3/7}\mu_\ast^{2/7}}$ \\[1ex]
& turbulent concentration & $\frac{2^{1/4}}{\pi\rho_s}\left(\frac{\Sigma m}{\alpha \sigma_{H_2}}\right)^{1/2}$ \\[1ex]
\hline\\[-1.5ex]
 Chondrule melting & splashing & $\frac{2\gamma}{GE\rho_p^2R_p^2}$\\[1ex]
 & impact melting & $\frac{0.07}{v_i}\left(\frac{\gamma R_i}{\rho_d}\right)^{1/2}$\\[1ex]
& droplet mergers & $\frac{\rho_d}{\rho_s}\Delta v t_m$\\[1ex]
\hline
\end{tabular}
\end{table}

\section{Summary and conclusions}
\label{Conclusion}

  I have investigated the origin of the weakly variable size of chondrules in chondrites. To that end, I have reviewed possible stages in the lifecycle of chondrules and their precursors, as broadly divided in: chondrule precursor growth, chondrule melting and sorting during aerodynamic transport. For these different processes
, I have expressed analytically the preferred chondrule size they would produce. Although I have strived to be as comprehensive as possible, I make no claim of completeness, nor should the formulas given be viewed as the definitive predictions of theories often still in development. Nonetheless, as they stand, they lend themselves to interesting evaluations against the meteoritical record. Indeed, although the processes envisioned were virtually all able to reproduce chondrule size for plausible values of the controlling parameters, few of them can, in their current state, account for the small variations of mean chondrule size among chondrite groups (excepting the outlying, and otherwise anomalous CH and CB groups). 

  I first examined the growth of potential chondrule precursors in the disk. Unless the chondrule- or chondrite-forming window was much narrower than suggested by age dating and petrographic evidence, fragmentation-limited growth of precursors would predict too wide variations to possibly be the size-determining process if the fragmentation velocity was constant, but could satisfy the constraint if the fragmentation velocity decreases with size as found by \citet{Beitzetal2011}, yielding a $(\Sigma/\alpha)^{0.51}$ dependence. Bouncing-limited growth, while still fraught with theoretical uncertainties, would quite robustly yield limited size variability (depending on the model and the disk parameters, dependences on surface densities would be to the 0.18-0.41 power), because of the opposite size dependences of the maximum sticking velocity and the collision velocities in the turbulent disk. It would be consistent with the comparable, albeit generally smaller, size of refractory inclusions
.  

  I then considered chondrule-melting processes themselves. Scenarios involving melt fragmentation like ``splash'' melting \citep{SandersScott2012,Asphaugetal2011} or impact melting \citep{Symesetal1998} were also found to predict too much variability (although they may account for the anomalously sized chondrules of the CH and CB groups)
. This would not \textit{per se} rule out these scenarios as chondrule-producing, but, unless the size predictions undergo revision, chondrule sizes would have to result from some postformational aerodynamic sorting (but see below). Based on empirical evidence from compound chondrules, droplet mergers were found to have a subordinate influence on the chondrule size distribution. The considerable compositional variability of chondrules would quite generally limit the size modifications that occurred during chondrule formation to less than a factor of a few. 

 I finally considered aerodynamic transport. Size sorting by radial transport would likely produce too much variability to qualify as the size-determining process. Turbulent concentration would predict likely suitably low variability (in $\left(\Sigma/\alpha\right)^{1/2}$), but lack of evidence of any accretion bias in carbonaceous chondrites argues against this and other synaccretional sorting having influenced chondrule size---although sorting may have occurred in chondrule-forming regions.

  Based on the above considerations, I infer that chondrule size was inherited from the precursors (similarly to previous suggestions by \citet{Sekiya1997,Zsometal2010}). Given the considerable uncertainty around growth by sticking \citep[e.g.][]{Ormeletal2007, Okuzumietal2009, Beitzetal2011,Windmarketal2012, Garaudetal2013,Kotheetal2013}, this preference relies less on confidence in the estimate of the outcome of growth by sticking herein 
than on a process of elimination of other competing mechanisms, because of size variability and/or independent cosmochemical arguments, and as such invites caution. If our inference were to hold true, however, chondrule-forming mechanisms would be constrained to involve simple melting of preexisting solid precursors, much like the conventional ``flash-heating'' picture, with little modification of size of the droplet (within a factor of a few). This would not necessarily imply that the cosmic setting of chondrule formation was ``nebular'' in a narrow (planetesimal-free) sense, especially given evidence of large pressures and/or solid densities there (\citet{Alexanderetal2008}; see e.g. \citet{Morrisetal2012}) but this would not fare well with the ``planetary'' scenarios envisioned above in their current state. Beyond the chondrule melting itself, chondrule size would also be a constraint for models of growth by sticking as indicating its outcome, and thence for models of further stages of primary accretion.

\section*{Acknowledgement}
I am grateful to Christopher Thompson and Yanqin Wu for interesting discussions on chondrule collisions. I also thank the three anonymous referees whose reviews significantly improved the discussion of growth by sticking and accretion bias, the interpretation of some meteoritical data, as well as the overall structure of the manuscript.


\begin{appendix}

\section{A model of compound chondrule formation}
\label{Compound model}

I propose here a simplified model of the collision between two droplets of equal radius ($a$) and the complete or incomplete relaxation of the newly formed object to sphericity.

  I restrict attention to temperatures where the plastic (viscous) behavior of chondrules wins over the elastic behavior, where collisions may result in mergers \citep{Cieslaetal2004}. (The transition temperature would presumably be near the glass transition temperature (around 1000 K; see \citet{AlexanderEbel2012}) or at any rate below 1400 K from the \citet{Connollyetal1994} experiments). Considering head-on collision trajectories for simplicity, the penetration length $x$ of the droplets into each other obeys Newton's second law:
\begin{equation}
m_r\frac{\mathrm{d}^2x}{\mathrm{d}t^2}=A_{\rm contact}(x)\sigma_{xx}
\end{equation}
with $m_r=(1/2)4\pi\rho_sa^3/3$ the reduced mass of the two colliding droplets, $A_{\rm contact}(x)\approx 2\pi a (x/2)$ the contact area between the two (for $x\ll a$) and $\sigma_{xx}$ the $xx$ component of the (viscous) stress tensor which I approximate as $-\eta_d(\mathrm{d}x/\mathrm{d}t)/a$ with $\eta_d$ the droplet viscosity
. Then, upon integrating the above, one finds that the two colliding droplets will come to rest ($\mathrm{d}x/\mathrm{d}t=0$) for
\begin{equation}
\label{penetration}
\frac{x}{a}=\frac{2}{\sqrt{3}}\left(\frac{\rho_sa\Delta v}{\eta_d}\right)^{1/2},
\end{equation}
with $\Delta v$ the initial relative velocity. For $\rho_sa=1\:\mathrm{kg/m^2}$ and $\Delta v\lesssim 1\:\mathrm{m/s}$, $\rho_sa\Delta v\lesssim 1\: \mathrm{Pa.s}$, much smaller than melt viscosities for T $\lesssim$ 1700-2000 K \citep{Giordanoetal2008}. Thus except perhaps for temperatures close to the liquidus, I have $x/a< 1$ (consistent with our assumption), so that the immediate result of the collision itself will be a bilobate object.

  Surface tension will however tend to restore a spherical shape. The relaxation timescale is \citep{Grossetal2013}
\begin{equation}
t_{\rm sph}=a\eta_d/\gamma. 
\end{equation}
with $\gamma$ the surface tension (taken to be 0.4 N/m). With decreasing temperature, the melt viscosity increases by orders of magnitude 
so that $t_{\rm sph}$ should eventually become longer than the cooling timescale at some temperature, below which collisions should yield compound chondrules (rather than larger spherical chondrules). For illustration, if I take the mesostasis composition of \citet{AlexanderEbel2012}, the viscosity model of \citet{Giordanoetal2008} and a cooling timescale of 1 h, I obtain a temperature of $\sim$1300 K. In this calculation, I have taken into account the viscosity enhancement of the droplet due to suspended crystals by adopting the \citet{Roscoe1952} formula:
\begin{equation}
\eta_d=\frac{\eta}{\left(1-c\right)^{5/2}}
\end{equation}
with $\eta$ the viscosity of the pure melt and $c$ the volume fraction of crystals taken to be 90 vol\%.

 \citet{Connollyetal1994} experimentally observed that collisions induced crystal nucleation at the interface in those droplets that were fully molten (such as those thought to solidify as nonporphyritic chondrules), and indeed most compound chondrules exhibit optical continuity at the junction \citep{Wassonetal1995}. For sufficient undercooling upon collision, crystal growth may have been sufficiently fast to prevent relaxation to sphericity at temperatures above the preceding threshold. 

  To quantify this, I adopt the crystal growth velocity $Y$ in the interface-controlled regime of \citet{Kirkpatrick1975}:
\begin{equation}
Y=\frac{fk_BT}{3\pi\eta a_0^2}\left(1-e^{-\mathcal{A}/k_BT}\right)
\end{equation}
with $a_0$ the molecular diameter (for which I adopt 0.27 nm---the edge of the silicate tetrahedron)
, $f$ the fraction of sites available for attachment (which I set to 1), and $\mathcal{A}$ the chemical affinity of the crystallization reaction
. I consider that the bilobate compound shape is frozen in if crystals can grow to a size $\gtrsim x$ within the relaxation timescale, that is if
\begin{equation}
\label{junction}
\frac{Yt_{\rm sph}}{x}=\frac{a}{x}\frac{fk_BT}{3\pi\gamma a_0^2}\left(1-e^{-\mathcal{A}/k_BT}\right)\gtrsim 1
\end{equation}
Approximating the affinity as $\mathcal{A}=L_c\left(T_L-T\right)/T_L$ \citep{Kirkpatrick1975} with $T_L$ the liquidus temperature and $L_c$ the latent heat of crystallization ($1.7\times 10^{-19}\:\mathrm{J}$ per silicate tetrahedron for pure forsterite \citep{Miuraetal2010}) and injecting equation (\ref{penetration}) yields, after a Taylor expansion of the exponential:
\begin{eqnarray}
\frac{Yt_{\rm sph}}{x} &=& \frac{1}{2\sqrt{3}\pi}\left(\frac{\eta}{\rho_sa\Delta v}\right)^{1/2}\frac{fL_c\left(T_L-T\right)}{\gamma a_0^2T_L}\nonumber\\
&=& 6\left(\frac{\eta}{10^2\:\mathrm{Pa.s}}\frac{1\:\mathrm{kg/m^2}}{\rho_sa}\frac{1\:\mathrm{m/s}}{\Delta v}\right)^{1/2}\frac{T_L-T}{T_L}
\end{eqnarray}
Taking the bulk type I chondrule composition of \citet{AlexanderEbel2012} to compute $\eta$ with the \citet{Giordanoetal2008} model, assuming $T_L=2000\:\mathrm{K}$, the above ratio reaches unity for the chosen normalizations at $T\approx 1500\:\rm K$, higher than the limiting temperature obtained from the sole cooling time constraint. 

  While this treatment is quite idealized and quantitative estimates should not be taken too seriously, this does show that nonporphyritic compound chondrules can form at higher temperatures than porphyritic ones because of nucleation upon collision, and thus explain their higher frequency.

\section{Epstein drag for nonspherical objects}
\label{Nonspherical Epstein}

\begin{figure}
\resizebox{\hsize}{!}{
\includegraphics{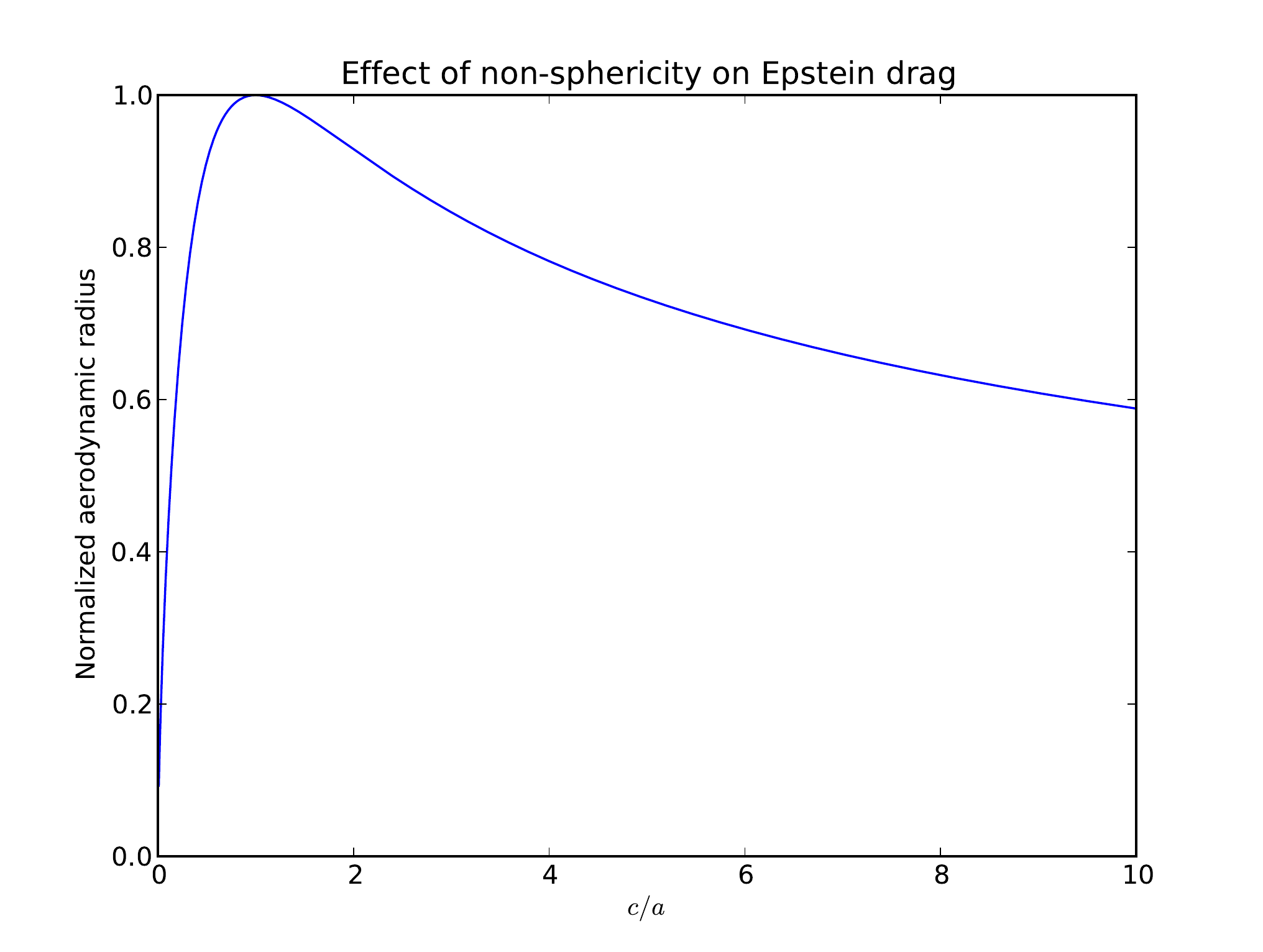}
}
\caption{Aerodynamic radius of a spheroid as a function of its axis ratio $c/a$ ($<1$ if oblate, $>1$ if prolate), normalized to the radius of the equal volume sphere. It is seen that a nonspherical spheroid is aerodynamically equivalent to a sphere of smaller volume than its own.}
\label{adrag spheroid}
\end{figure}

In this appendix, I provide the expression of the drag force $\mathbf{F}$ of the gas on a solid object much smaller than the molecular mean free path. The calculation, which ignores molecules reflected toward the object itself, and is therefore strictly valid only for convex shapes\footnote{Unlike fractal aggregates \citep{Okuzumietal2012} for which drag laws have yet to be ascertained, although for fractal dimensions $\leq 2$, the mean free path of molecules (with respect to collisions with monomers) in the aggregate would be longer than its size. For nonfractal but invaginated objects, I suggest taking the convex enveloppe as the effective surface to estimate drag.}, follows the same lines than that of \citet{Dahneke1973} although the force is here expressed in a coordinate-free vector fashion. I thus only quote the result:
\begin{equation}
\mathbf{F}=\frac{\rho v_T}{4}\left(A\mathbf{u}+\left(1+\beta\right)\int \mathbf{u}\cdot\mathbf{n}\:\mathrm{d}\mathbf{A}\right),
\label{drag general}
\end{equation}
with $v_T=\sqrt{8/\pi}c_s$, $\mathrm{d}\mathbf{A}$ 
 the surface element vector (pointing outward) and $\mathbf{n}$ the corresponding unit vector, $A$ the integrated (\textit{not} projected) surface area, $\mathbf{u}$ the mean velocity of the gas relative to the solid. The parameter $\beta$ is $9\pi/16$ for a perfectly non-conducting solid and $\pi\sqrt{T_p/T}/2$ for a perfect conductor (with $T_p$ the temperature of the particle)\footnote{For specular reflection, I have $\mathbf{F} = \rho v_T\int \mathbf{u}\cdot\mathbf{n}\mathrm{d}\mathbf{A}$, the orientation averaging of which yields the aerodynamic radius given by equation (\ref{aerodynamic radius}) if I set $\beta=0$.}. The formula assumes no rotation of the solid but the contribution of rotation would be zero if it possesses a center of symmetry (in which case the only contribution to the torque would be a braking of the rotation). If the formula is averaged over all possible orientations of the solids relative to the flow, I obtain:
\begin{equation}
\langle\mathbf{F}\rangle = \frac{\rho v_T}{3}A\left(1+\frac{\beta}{4}\right)\mathbf{u},
\end{equation}
hence a stopping time which can be cast in the usual form
\begin{equation}
\tau\equiv\frac{m_pu}{F}\equiv\frac{\rho_sa_{\rm drag}}{\rho v_T\left(1+\beta/4\right)},
\end{equation}
(where the $(1+\beta/4)$ correction may generally be ignored), if the ``aerodynamic radius'' is defined as:
\begin{equation}
a_{\rm drag}=\frac{3V}{A},
\label{aerodynamic radius}
\end{equation}
where $m_p$ and $V=m_p/\rho_s$ are the mass and volume of the solid, respectively. This corresponds to the actual radius in the case of a sphere. For a spheroid of equatorial and polar radii $a$ and $c$, respectively:
\begin{equation}
a_{\rm drag}=\frac{2c}{1+(c/a)f(a/c)},
\end{equation}
with
\begin{equation}
 f(x)=
\left\{\begin{array}{rrr}
\frac{\mathrm{arccos}x}{\sqrt{1-x^2}}\:\mathrm{if}\:x<1\\
\frac{\mathrm{argcosh}x}{\sqrt{x^2-1}}\:\mathrm{if}\:x>1\\
1\:\mathrm{if}\:x=1
\end{array}\right.
\end{equation}
In figure \ref{adrag spheroid}, $a_{\rm drag}$ is plotted normalized to the radius of the equal volume sphere. One sees that the spherical aerodynamic equivalent of a spheroid is smaller than the latter. This is a general consequence of the isoperimetric inequality $36\pi V^2/A^3\leq 1$.  Thus, irregularly shaped inclusions aerodynamically equivalent to (spherical) chondrules should be \textit{bigger} than those.

\end{appendix}

\bibliographystyle{natbib}
\bibliography{bibliography}

\end{document}